\documentclass[conference]{IEEEtran}
\IEEEoverridecommandlockouts
\usepackage{cite}
\usepackage{amsmath,amssymb,amsfonts}
\usepackage{algorithmic}
\usepackage{graphicx}
\usepackage{textcomp}

\usepackage{url}
\usepackage{mathtools}
\usepackage{flushend}
\usepackage{color}
\usepackage{epstopdf}
\usepackage{subfig}
\usepackage{float}
\usepackage{enumerate}
\usepackage{multirow}
\usepackage{array}
\usepackage{pifont}
\usepackage[font={footnotesize}]{caption}
\usepackage[table,xcdraw]{xcolor}
\usepackage{wasysym}
\usepackage{xspace}
\usepackage{listings}
\usepackage{amsmath}

\definecolor{dkgreen}{rgb}{0,0.6,0}
\definecolor{gray}{rgb}{0.5,0.5,0.5}
\definecolor{mauve}{rgb}{0.58,0,0.82}

\newcommand{\xeg}{e.g.\@\xspace}

\newcommand{\xie}{i.e.\@\xspace}

\def\longitem[#1]{\item[] {\normalfont\bfseries #1}}

\def\BibTeX{{\rm B\kern-.05em{\sc i\kern-.025em b}\kern-.08em
    T\kern-.1667em\lower.7ex\hbox{E}\kern-.125emX}}
\begin{document}

\title{Colocating Real-time Storage and Processing: An Analysis of Pull-based versus Push-based Streaming}

\author{\IEEEauthorblockN{Ovidiu-Cristian Marcu, Pascal Bouvry}
\IEEEauthorblockA{\textit{University of Luxembourg, Luxembourg} \\
ovidiu-cristian.marcu@uni.lu, pascal.bouvry@uni.lu}
}

\maketitle

\begin{abstract}

Real-time Big Data architectures evolved into specialized layers for handling data streams' ingestion, storage, and processing over the past decade. Layered streaming architectures integrate pull-based read and push-based write RPC mechanisms implemented by stream ingestion/storage systems. In addition, stream processing engines expose source/sink interfaces, allowing them to decouple these systems easily. However, open-source streaming engines leverage workflow sources implemented through a pull-based approach, continuously issuing read RPCs towards the stream ingestion/storage, effectively competing with write RPCs. This paper proposes a unified streaming architecture that leverages push-based and/or pull-based source implementations for integrating ingestion/storage and processing engines that can reduce processing latency and increase system read and write throughput while making room for higher ingestion. We implement a novel push-based streaming source by replacing continuous pull-based RPCs with one single RPC and shared memory (storage and processing handle streaming data through pointers to shared objects). To this end, we conduct an experimental analysis of pull-based versus push-based design alternatives of the streaming source reader while considering a set of stream benchmarks and microbenchmarks and discuss the advantages of both approaches.

\end{abstract}


\section{Introduction}

Fast data storage and streaming architectures are deployed intensively in both Cloud \cite{dataflow,Hazelcast} and Fog architectures \cite{fogStreaming}. Real-time data-intensive processing can require very low-latency \cite{granular}. E.g., implementing sensitive information detection with the NVIDIA Morpheus AI framework enables cybersecurity developers to create optimized AI pipelines for filtering and processing large volumes of real-time data \cite{morpheus}. Moreover, fast data processing exquisites low-latency and high-throughput data access to streams of logs, e.g., daily processing terabytes of logs from tens of billions of events at CERN accelerator logging service \cite{nextCernALS, nxcalsArchitecture}.

\begin{figure} [t]
\centerline{\includegraphics[width=9cm]{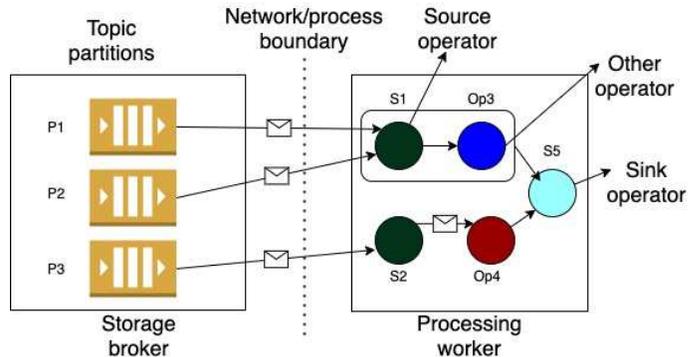}}

\caption[Integration of storage brokers and streaming engines through source operators]
{Pull-based real-time stream sources (e.g., implementing Remote Procedure Calls or RPCs) continuously issue RPCs to obtain next record chunks of stream. Current streaming architectures are composed of messaging brokers and stream processing engines and are decoupled through sources and sinks. Each broker handles a set of topic partitions. Each processing worker can deploy three types of operators: sources, sinks, and other processing operators (e.g., map, filter). The source operator S1 pulls messages from partitions P1 and P2, while the source operator S2 pulls messages from partition P3. The source S1 and operator Op3 can be chained and potentially executed in the same task or processing slot. In contrast, the source S2 and operator Op4 are deployed for execution separately, e.g., communicating through shared queues. Finally, the sink operator S5 accumulates stream records from operators Op3 and Op4 and is responsible for writing them out.}

\label{fig:source}
\end{figure}

Layered streaming architectures integrate through reading and writing RPC APIs implemented by stream storage systems and through source and sink operators' interfaces implemented by stream processing engines, allowing for efficiently decoupling the two systems. The streaming source operator pulls data from the storage brokers' assigned topic partitions. However, since data streams are unpredictable, stream workflow pipelines \cite{trisk} can trigger an updated pipeline execution, including repartitioning input stream partitions to stream sources. Therefore, the role of the streaming source is critical as it participates in an optimized dynamic streaming workflow. E.g., as illustrated in Figure \ref{fig:source}, sources pipelined with other operators can help in reducing communication and de/serialization), handling the rate of the stream to manage backpressure, and may be responsible for discovering new partitions (e.g., dynamic partitioning in KerA \cite{keraIcdcs}).

While the streaming source is rich functionally, it can also be the source of bottlenecks and overall reduced application performance. For example, the source parallelism (i.e., how many deployed source tasks for consuming stream partitions) impacts resource usage and processing latency/throughput (triggering reconfigurations). Therefore, for streaming source design and implementation, streaming architects consider a pull-based approach design since it simplifies implementation and gives complete control to stream system developers, allowing them to decouple layered streaming architectures. This choice is opposed to monolithic architectures \cite{pangea} that have the opportunity to more efficiently optimize data-related tasks. However, a push-based approach for integrating streaming sources with real-time storage can bring essential improvements if carefully architected. E.g., since stream data could be pushed by the storage broker as soon as it is available, it can effectively reduce latency and increase system throughput. Furthermore, when the network is the bottleneck, one can colocate storage and processing, simplifying the push-based source implementation. One of the challenges in integrating a push-based streaming source approach is keeping control of the stream consumption, which is easier to implement when choosing a pull-based design (e.g., as implemented in state-of-the-art streaming engines like Apache Spark or Apache Flink). 

Our challenge is then \textbf{how to design and implement a push-based streaming source strategy to efficiently and functionally integrate real-time storage and streaming engines} while keeping the advantages of a pull-based approach. Towards this goal and efficiently optimizing streaming throughput while reducing processing latency, this paper introduces a push-based streaming design to integrate real-time storage and processing engines. Furthermore, this paper explores the pull-based and push-based approaches for the stream source operator design through extensive empirical evaluations. We make the following contributions:

\begin{itemize}
    \item We describe challenges introduced by layered streaming storage and processing architectures.

    \item We design a unified real-time storage and streaming architecture by introducing a push-based source protocol that maintains pull-based approach properties such as backpressure.

    \item We implement pull-based and push-based stream sources as integration between KerA \footnote{KerA's source-based integration code will be available at: https://gitlab.uni.lu/omarcu/zettastreams.} real-time storage system and Apache Flink.

    \item We evaluate the KerA-Flink integration over a set of benchmarks. We empirically show that the push-based source approach can be competitive with a pull-based design while requiring reduced resources. Furthermore, when storage resources are constrained, our results sustain the shared wisdom of processing data locally first despite high-performance networks \cite{reslowNetworks}: the push-based approach can be up to 2x more performant compared to a pull-based design.

\end{itemize}

\section{Background and Motivation}

\subsection{Streaming Architectures and Locality Related Work}

Unbounded stream topics are infinite data sets that accumulate stream records from multiple producers (and data sources) while having numerous consumers subscribing to these topics (e.g., Apache Kafka \cite{kafkapaper}). Decoupling producers and consumers through message brokers can help applications through simplified architectures: e.g., availability and durability of data streams are managed separately from processing engines. This locality-poor design is preferred over monolithic architectures by state-of-the-art open-source streaming architectures.

State-of-the-art Big Data frameworks that implement the MapReduce paradigm \cite{MapReduce} are known to implement data locality optimizations. General Big Data architectures can thus efficiently co-locate map and reduce tasks with input data, effectively reducing the network overhead and thus increasing application throughput. However, they are not optimized for low-latency streaming scenarios. 

Moreover, machine learning optimizations \cite{sparkMemoryCoLoc} that co-locate memory-aware tasks can further improve system throughput, providing system optimizations orthogonal to ours. Finally, user-level thread implementations such as Arachne \cite{arachne} and core-aware scheduling techniques like Shenango, Caladan \cite{shenango, caladan} can further optimize co-located latency-sensitive stream storage and analytics systems. 

Finally, it is well known that message brokers, \xeg, Apache Kafka \cite{apachekafka}, Apache Pulsar \cite{pulsar}, Distributedlog \cite{distributedlog}, Pravega \cite{pravega}, or KerA \cite{keraIcdcs}, can contribute to higher latencies in streaming pipelines \cite{flinkKafka}. Indeed, none of these open-source storage systems implement locality and thus force streaming engines to implement a pull-based approach for consuming data streams. Consistent state management in stream processing engines is difficult \cite{stateFlink} and depends on real-time storage brokers to provide indexed, durable dataflow sources. Therefore, source design is critical to the fault-tolerant streaming pipeline and potentially a performance issue.

We believe our work is at the intersection between monolithic architectures \cite{pangea} (that have the opportunity to more efficiently optimize data-related tasks) and decoupled layered streaming architectures that do not benefit from data locality optimizations. As far as we know, this paper presents for the first time an analysis of push-based versus pull-based streaming source deployments in Fast Data architectures.

\subsection{Pull-based versus Push-based Streaming Challenges}

Stream processing engines implement consumers through multi-threaded readers and call these integration stream sources. For streaming architectures to scale, stream topics get partitioned (e.g., static partitioning in Kafka or dynamic partitioning in KerA \cite{keraIcdcs}). Source readers are assigned one or multiple topic partitions, while each partition is associated with an offset from which to start reading. Therefore, source readers have various roles: (1) consume stream tuples from their associated partitions through either \emph{pull-based} or \emph{push-based} RPC approaches, (2) deserialize messages and make them available to pipelined stream tasks through queues, (3) participate in the fault-tolerant streaming pipeline, e.g., caching stream tuples, emitting watermarks \cite{watermarks}, re-consume stream tuples from older partition offsets.

As illustrated in Figure \ref{fig:source}, state-of-the-art stream processing engines implement pull-based source readers to manage backpressure \cite{backpressure} easily—this situation in which slow stream operators tend to bottleneck streaming pipelines, rapidly filling sources' queues. In addition, the pull-based approach brings the advantage of completely decoupling processing from storage while giving architects more flexibility for handling scenarios with diverse performance requirements. A pull-based source reader works as follows: it waits no more than a specific timeout before issuing RPCs to pull (up to a particular batch size) more messages from stream partitions. It is difficult to tune these source parameters (timeout, RPC batch size) for every workload. However, high-throughput workloads may benefit more from a pull-based approach than low-latency scenarios, except when the network is the bottleneck.

One crucial question is how much data these sources have to pull from storage brokers and how often these pull-based RPCs should be issued to respond to various application requirements. Consequently, a push-based approach can quickly solve these issues by pushing the following available messages to the streaming source as soon as more stream messages are available. However, a push-based source reader is more difficult to deploy since coupling storage brokers and processing engines can bring back issues solved by the pull-based approach (e.g., backpressure, scalability).

Since stream topics get partitioned for scalability, streaming sources should dynamically consume these partitions. However, the number of partitions is unknown at runtime. Therefore, we should configure the number of sources (also called source parallelism) based on the application's throughput and latency requirements. In addition to these parameters, the streaming workflow chain optimizations are another source of configuration complexity; as illustrated in Figure \ref{fig:source}, streaming sources, sinks, and other operators can be chained for execution to optimize buffering and thus throughput. Assume the streaming operator is a map or a filter: a pipelined operator deployment can quickly reduce the size and volume of stream messages, effectively reducing latency and increasing throughput).

When high-performance networking (e.g., Infiniband) can be leveraged \cite{hpmsgbrokers}, streaming latencies can be highly reduced while larger volumes of data streams can be acquired and processed. However, a pull-based source approach can contribute to inefficient streaming architectures. Broker architecture RPCs are handled by a multi-threaded dispatcher-workers architecture (e.g., see RAMCloud \cite{ramcloud}). Sources' RPCs compete with producers, while the broker's dispatcher thread can quickly become a bottleneck \cite{arachne} in low-latency scenarios.

Another issue is source scheduling. Co-locating source operators with their partitions (thus brokers) may be more efficient when the network is scarce. However, when source operators get chained with other streaming operators, CPU usage is an additional factor to be considered in optimizing streaming workflows.

Given the issues mentioned above, our intuition is that we can optimize streaming architectures (that decouple message brokers from processing engines) by considering push-based streaming sources co-located with storage brokers whenever possible. However, since multiple parameters contribute to the source design complexity, let us define our problem statement further.

\section{Problem Statement}

We consider a producer-consumer streaming model where Np producers append events in parallel to Ns independent stream partitions. At the same time, Nc consumers can sequentially read at any offset from associated partitions, with one partition exclusively processed by one consumer. Consumers are part of a streaming workflow and further collaborate with other streaming operators and sinks through queues. Thus, streaming consumers have a limited cache for storing stream data before pushing it to the other operators. Since producers and consumers compete on storage resources, our goal is twofold. First, we want to maximize the overall throughput of appends (producers) with concurrent reads. Second, we want to maximize the overall throughput of reads (consumers) while reducing processing latency. At scale, this is challenging since consumers do not know at deploy time how much data to consume from available stream partitions.

Consumers can implement two strategies for consuming stream data. State-of-the-art streaming engines employ a \emph{pull-based} approach in which consumers issue RPCs to pull data from assigned partitions. Each consumer RPC can consume up to a defined CS chunk size for each partition. However, tuning this parameter to efficiently optimize storage resources while giving room to producers and other consumers is difficult. Another approach is to leverage a \emph{push-based} approach in which the storage broker is responsible for pushing available stream data as they arrive. However, since the processing engine is losing source control, it is challenging to ensure a backpressure mechanism with a naive push-based approach. Moreover, since multiple sources can consume data from partitions of one storage broker node, we are interested in optimizing the (shared) resources dedicated to source reader management.

Our challenge is then \textbf{how to design and implement a push-based streaming source strategy to efficiently and functionally integrate real-time storage and streaming engines} while keeping the advantages of a pull-based approach. Towards this goal and efficiently optimizing streaming throughput while reducing processing latency, let us introduce next a push-based streaming design that integrates real-time storage and processing engines and describe our implementation.

\begin{figure}[t]
\centering

\includegraphics[width=8.5cm]{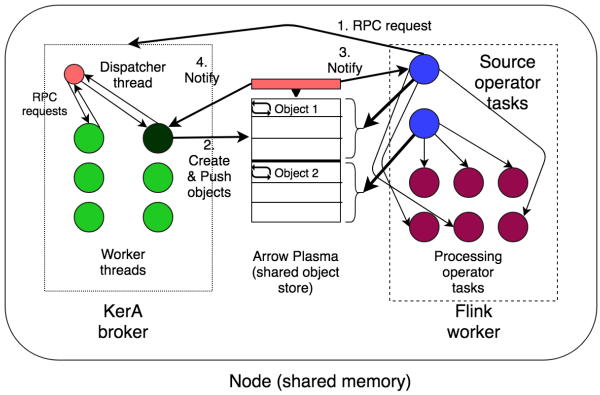}

\caption[Data locality architecture through shared in-memory object store]
{Unified real-time storage and processing architecture with push-based consumers through shared in-memory object store. On the same
node live three processes: the streaming broker (KerA), the processing worker
(Flink) and the shared object store (Arrow Plasma). Source tasks coordinate to
launch one RPC request (step 1). The worker thread is responsible to fill
shared objects with next stream data (step 2). Source tasks are notified for
object updates (step 3) and process new stream data. The worker thread is
notified (step 4) after each source processed all objects, so a new
`iteration' for that source can be started. This flow executes continuously.}
\label{fig:locality}
\end{figure}

\section{Unified Real-time Storage and Processing Architecture: Our Design and Implementation}

\subsection{Background}

\textbf{Streaming Storage Broker Architecture}. A streaming storage architecture contains one coordinator that manages cluster metadata, recovery, and initial communication with clients and a layer of \emph{B} brokers that serve producers and consumers of data streams. As illustrated in Figure \ref{fig:locality}, a broker is configured with one dispatcher thread (one CPU core) polling the network and responsible for serving RPC requests and multiple working threads that do the actual writes and reads to data stream partitions (for more details check \cite{vlogKerA}).

\textbf{Streaming Processing Worker Architecture}. A streaming processing architecture contains one master and a layer of \emph{N} workers. E.g., in Apache Flink, each worker implements a JVM process that can host multiple slots (a slot can have one core). In addition, sources, sinks, and other operators are deployed at runtime on worker slots (for more details, check \cite{flink, spark}). 

\emph{Big Data streaming architectures are typically designed to scale to
a large number of simultaneous pull-based consumers that enable processing for millions of records per second, \cite{drizzle,streambox}. Thus, the weak link of the three-stage pipeline is the ingestion phase: it needs to acquire
records with high throughput from the producers, serve the consumers with a
high throughput, scale to a large number of producers and consumers, and
minimize the write latency of the producers and, respectively, the read
latency of the consumers to facilitate low end-to-end latency \cite{keraIcdcs}.} 

Since producers and consumers communicate with message brokers through RPCs, there is
inevitably interference between these operations, leading to increased
processing times. Moreover, since consumers (\xie, source operators) depend on
the networking infrastructure, its characteristics can limit the read
throughput and increase the end-to-end read latency. One approach is to
co-locate processing workers (source and other operators) with brokers
managing stream partitions. However, this is not enough since the competition 
between producers and consumers remains the same. To tackle this challenge, we propose
a push-based approach for consuming streams.

\subsection{Our Push-based Streaming Design for Real-time Sources}

\textbf{Our architecture for co-locating real-time storage and processing engines.} As illustrated in Figure \ref{fig:locality}, we deploy a storage broker and one or multiple processing workers on a multi-core node. Storage and processing engines communicate by default through pull-based RPCs. We propose an architectural extension based on shared memory techniques that allow streaming source operators to leverage locality by proper in-memory support and to access stream data at lower latencies and potentially higher throughput.

We propose to leverage a shared buffer between (push-based) streaming sources and storage brokers to provide backpressure support to streaming engines and to allow for transparent integration with various streaming storage and processing engines. Our design principle is that various storage engines commonly implement push-based APIs, while our shared memory technique should allow for efficiently using shared storage resources by streaming engines. Furthermore, the size of the shared (partitioned) in-memory buffer and the number of dedicated storage resources should be determined dynamically at runtime based on application needs.

\textbf{Our push-based streaming design and implementation.} Let us describe our example illustrated in Figure \ref{fig:locality} that provides locality support for streaming operations. In this case, two push-based streaming source tasks are scheduled on one processing worker. Each source task implements one thread that can build a push-based RPC to initiate requesting data (Step 1. RPC request). At runtime, only one of the two sources will issue the push-based RPC (e.g., based on the smallest of the source tasks' identifiers). This RPC request contains initial partition offsets used by sources to consume the next chunks of records. Alternatively, the storage broker can assign local partitions and build consumer offsets. The storage handles the push-based RPC request by assigning a worker thread responsible for creating and pushing the next chunks of data (Step 2. Create and push objects) associated with consumers' partition offsets.

Replacing pull-based consumer RPCs with one dedicated worker thread that continuously
pushes records to consumers through shared memory helps reduce the
interference with producer requests. These improvements can be thought as
similar to optimizations brought by techniques such as Tailwind \cite{tailwind}: less
competition on dispatcher and worker threads leaves more CPU space for
executing producers' ingestion and backup RPC requests which translates into more ingestion and processing throughput. However, we should be careful when choosing how many sources can share a dedicated broker thread based on throughput and latency requirements.

A shared partitioned memory object store sits between the storage broker and the processing worker. We need to identify each in-memory chunk of data by its object identifier to communicate between broker and streaming worker, reusing them efficiently. Our shared partitioned object store is leveraged by all local source tasks of a worker as follows. We partition the object-based store into objects that give access through a pointer to their memory. Both broker and sources use the object pointer based on notifications. Local streaming sources synchronize so that only one
RPC request is sent to the broker, and one worker thread implements the request
as a normal consumer source (managing offsets internally). The broker then
pushes chunks through shared objects and notifies streaming sources (Step 3. Notify sources) when each object is updated (object buffers are reused). Once a streaming source processes its
objects, it notifies the broker (Step 4. Notify broker) to push more chunks by reusing them. 

We implement a shared-memory object-based store based on Apache Arrow Plasma \cite{ray}, a framework that allows the creation of in-memory buffers (named objects) and their manipulation through shared pointers. Our push-based RPC is implemented on the KerA storage engine while we integrate with Apache Flink. This work consists of about 4K lines of C++ code for client and server-side implementations and 2K lines of Java code for integrating with Apache Flink. Future integration with various streaming engines can reuse our streaming connector.

\section{Evaluation}

While existing streaming benchmark efforts \cite{Theodolite,benchdsp,RIoTBench} target the scalability or performance metrics of the stream processing engines, our goal is to evaluate and understand the impact of streaming sources on performance. Therefore, we compare stream source deployments that leverage the pull-based and push-based approaches in real-time layered streaming architectures that decouple storage brokers from processing engines. We chose both a set of stream applications for benchmarking stream processing systems and a set of microbenchmarks to help us understand a fine-grained tuning of the streaming sources.

Our streaming architectural implementation used for evaluation is composed of KerA \cite{vlogKerA}, a high-performance replicated message broker, and Apache Flink \cite{flink}, a scale-out Java-based streaming engine. We choose KerA since it delivers better throughput than Apache Kafka and its architecture allows leveraging both commodity and high-end networks like Infiniband. Although scale-up C++ stream processing alternatives \cite{modernStreaming} deliver orders of magnitude better performance than Java-based systems like Apache Flink, application development is not easy. Being widely adopted by industry and academia, we choose Apache Flink since it also uses a tuple-at-a-time processing model that is more appropriate for real-time processing on a scale-out cluster. Furthermore, we run our evaluation on multi-core nodes connected with high-end networking like Infiniband to be as relevant as possible to future cluster deployments.

\subsection{Experimental Setup and Parameter Configuration}

We execute our benchmarks on the Aion cluster \footnote{more details at https://hpc.uni.lu/infrastructure/supercomputers the Aion section} by deploying Singularity containers over Aion regular nodes. Aion nodes have 2 AMD Epyc ROME 7H12 CPU 64 cores, each with 256 GB of RAM, interconnected through Infiniband 100Gb/s network through Slurm jobs. Given our cluster configuration, we avoid the networking communication becoming a bottleneck. We choose multi-core nodes to co-locate storage and processing in multiple configurations easily. A set of producers are deployed separately from the streaming architecture. We provide producers' configuration in the evaluation subsection. The KerA message broker is configured with up to 16 worker cores/threads while the partition's segment size is fixed to 8 MiB. We use Apache Flink version 1.13.2. As opposed to \cite{modernStreaming},  data is ingested and consumed in real-time to evaluate streaming architectures in real deployments properly.

Table \ref{tab:parameters} lists the use of the most important parameters by each benchmark. We configure several producers \emph{Np} (values={1,2,4,8}) that produce and push chunks of data of a stream having \emph{Ns} partitions. Producers and pull-based consumers are multi-threaded and are configured similarly to \cite{vlogKerA}. The number of consumers \emph{Nc} (values={1,2,4,8}) is chosen such that for each partition, there is one consumer; each partition is consumed exclusively by its associated consumer. Each producer issues one synchronous RPC having one chunk of \emph{CS} size (values={1,2,4,8,16,32,64,128} KiB) for each partition of a broker, having in total \emph{ReqS} size. Each chunk can contain multiple records of configurable \emph{RecS} size for the synthetic workloads. We configure producers to read and ingest Wikipedia files in chunks having records of 2 KiB. Flink workers correspond to the number of Flink slots \emph{NFs} (values={8,16}) and are installed on the same Singularity instance where the broker lives. When a backup is configured (\emph{Replication} is two), it lives on a separate Aion node. Pull-based consumers and producers continuously issue synchronouss RPCs. Our push-based consumers leverage one dedicated thread and consume shared objects as described in the implementation section.

\begin{table}[t]
\centering
\begin{tabular}{|c|c|}
\hline
\emph{Np} & Number of producers \\ \hline
\emph{Nc} & Number of consumers, the sourceParallelism \\ \hline
\emph{Nmap} & Number of application mappers, the mapParallelism \\ \hline
\emph{Ns} & Number of stream partitions \\ \hline
\emph{CS} & Chunk size \\ \hline
\emph{ReqS} & Request size, one chunk for each partition \\ \hline
\emph{RecS} & Record size \\ \hline
\emph{Replication} & Partition replication \\ \hline
\emph{NBc} & Number of KerA broker working cores \\ \hline
\emph{NFs} & Number of Flink processing slots \\ \hline
\end{tabular}
\caption{Parameters used in benchmarks.} 
\label{tab:parameters}

\end{table}

\subsection{Benchmarks}

This section presents a set of benchmarks we devise to understand the performance differences between a pull-based and a push-based strategy for streaming consumers. In all our benchmarks, we measure cluster throughput (in millions of tuples per second) by aggregating the throughput of every producer/consumer in each second.

\begin{itemize}
\item \textbf{Synthetic benchmarks.} We have selected two benchmarks that leverage synthetic data. Producers are configured to push data through RPCs over a stream configured to have multiple partitions. The first benchmark implements a simple pass-over data, iterating over each record of partitions' chunks while counting the number of records per second for each source configured to consume records produced and consumed concurrently. This benchmark is relevant to use cases that transfer or duplicate partitioned datasets. However, the source is only consuming and counting records to understand the maximum throughput that can be obtained in real-time. The second benchmark implements a filter function over each record, adding to the CPU consumption, and therefore we expect throughput to be slightly reduced compared to the first benchmark. The filter (or grep operation) is a representative workload used in several real-life applications, either scientific (\emph{e.g.} indexing the monitoring data at the LHC \cite{lhc}) or Internet-based (\emph{e.g.} search at Google, Amazon \cite{googlealgorithms}).

\item \textbf{Wikipedia benchmarks.} We have opted for two benchmarks implementing the Word Count with and without sliding windows (window size equals five seconds, sliding each second). Similarly, the source and word count mappers are configured with different parallelism, although some tasks are pipelined at deployment time. The Word Count benchmarks are more CPU intensive, and we are interested in understanding the source parallelism's impact on aggregated throughput.
\end{itemize}

\begin{lstlisting}[caption=Synthetic workloads]
//count and filter tuples

FlinkKeraConsumer keraConsumer = 
new FlinkKeraConsumer(topic,schema,props);

DataStream<Tuple2<byte[],byte[]>> 
cons=env.addSource(keraConsumer)
.setParallelism(sourceParallelism)
.flatMap(
new RTLogger<Tuple2<byte[],byte[]>>())
.setParallelism(mapParallelism)
.write("output.txt").setParallelism(1);
\end{lstlisting}

\begin{lstlisting}[caption=Wikipedia workloads]
//streaming word count
DataStream<Tuple2<byte[],byte[]>>
cons=env.addSource(keraConsumer)
.setParallelism(sourceParallelism);

DataStream<Tuple2<String,Integer>> 
counts = cons.flatMap(new Tokenizer())
.setParallelism(mapParallelism)
.keyBy(value -> value.f0).sum(1)
.flatMap(
new RTLogger<Tuple2<String,Integer>>())
.setParallelism(mapParallelism)
.write("output.txt").setParallelism(1);

//streaming windowed word count
DataStream<Tuple2<String,Integer>> 
wCounts=cons.flatMap(new Tokenizer())
.setParallelism(mapParallelism)
.keyBy(value -> value.f0)
.countWindow(windowSize, slideSize)
.sum(1).flatMap(
new RTLogger<Tuple2<String,Integer>>())
.setParallelism(mapParallelism)
.write("output.txt").setParallelism(1);
\end{lstlisting}

Let us describe these benchmark applications to evidence the two parallelism configurations we benchmark, separately for the source and the mapper operators doing the count, filter, and word count work. As illustrated in Listing 1, each DataStream of tuples will configure the stream source with the sourceParallelism, while the flatMap operators are configured with a higher mapParallelism. Finally, the sink writeAsText operator will log every second the throughput computed by the flatMap function implemented by RTLogger for counting. Similarly, the filter operation uses a RichFilterThroughputLogger function that applies a filter operation on the string represented by the byte array value of each tuple). We illustrate the word count benchmark applications in the Listing 2 - source and map parallelism are applied similarly. We summarize our benchmark evaluations and associated application operators in Table \ref{tab:experiments}.

\begin{figure}[t]
\centering

\includegraphics[width=8.5cm, height=4.3cm]{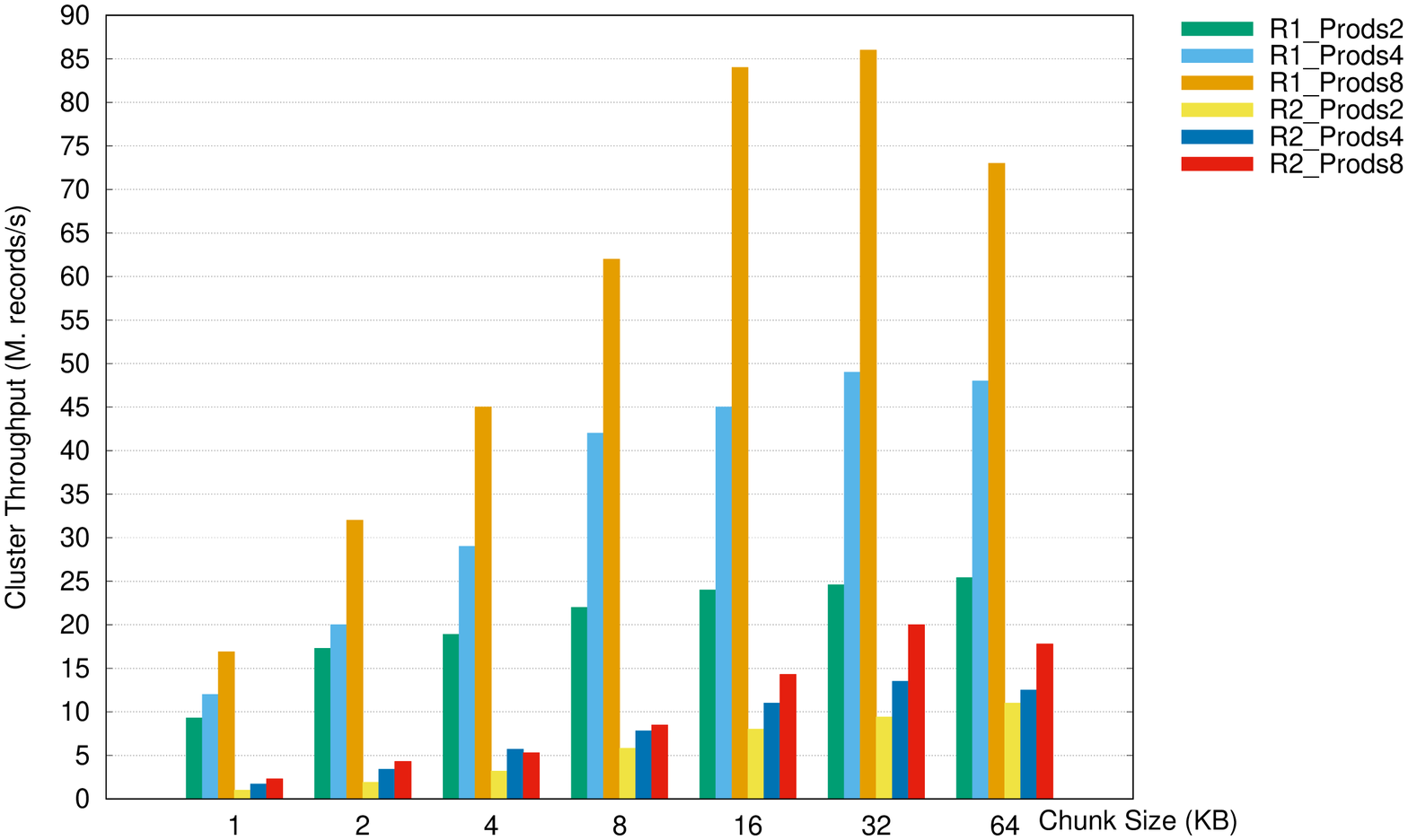}
\caption{Ingestion benchmark with 2, 4, and 8 concurrent producers, record size 100 Bytes, no key, one stream with 8 partitions (similar results for 16 partitions). While scaling the number of producers we increase the partition chunk size. Each vertical line represents one experiment and plots the aggregated producers' throughput records per second. R1Prods2 corresponds to two producers writing chunks of data that are kept in one single copy in memory by the storage broker, while R2Prods8 corresponds to eigth producers with replication factor two.}
\label{fig:expOnlyProds}
\end{figure}

\subsection{Evaluation: Results and Discussion}

To understand previous parameters' impact on performance and quantify the differences between a pull-based strategy and a push-based strategy for streaming consumers, we run a set of experimental benchmarks that work as follows. We run each experiment for 60 to 180 seconds while we collect producer and consumer throughput metrics (records/tuples every second). We plot 50-percentile aggregated throughput per second for each experiment (i.e., summing producer and consumer throughputs), and we compare various configurations to understand the trade-offs introduced by the push-based strategy for streaming consumers. Our goal is to understand \emph{who of the push-based and respectively pull-based streaming strategies is more performant and what the trade-offs are in terms of configurations.}

\begin{table}[t]
\centering
\begin{tabular}{|c|c|c|c|c|}
\hline
\textbf{Benchmarks Pull versus Push} & Filter & Count & Map & KeyBy  \\ \hline
\emph{Count Broker 16 cores Fig.4} & & \checkmark & \checkmark & \\ \hline
\emph{Filter 8 partitions Fig.5} & \checkmark & \checkmark & \checkmark & \\ \hline
\emph{Filter 4 partitions Fig.6} & \checkmark & \checkmark & \checkmark & \\ \hline
\emph{Filter Broker 4 cores Fig.7} & \checkmark & \checkmark & \checkmark & \\ \hline
\emph{Small Chunks Broker 8 cores Fig.8} & & \checkmark & \checkmark & \\ \hline
\emph{Windowed Word Count Fig.9} & & \checkmark & \checkmark & \checkmark  \\ \hline

\end{tabular}
\caption{Benchmarks and related operators.} 
\label{tab:experiments}

\end{table}

\begin{figure*}[t]
\centering

\includegraphics[width=5.4cm, height=2.7cm]{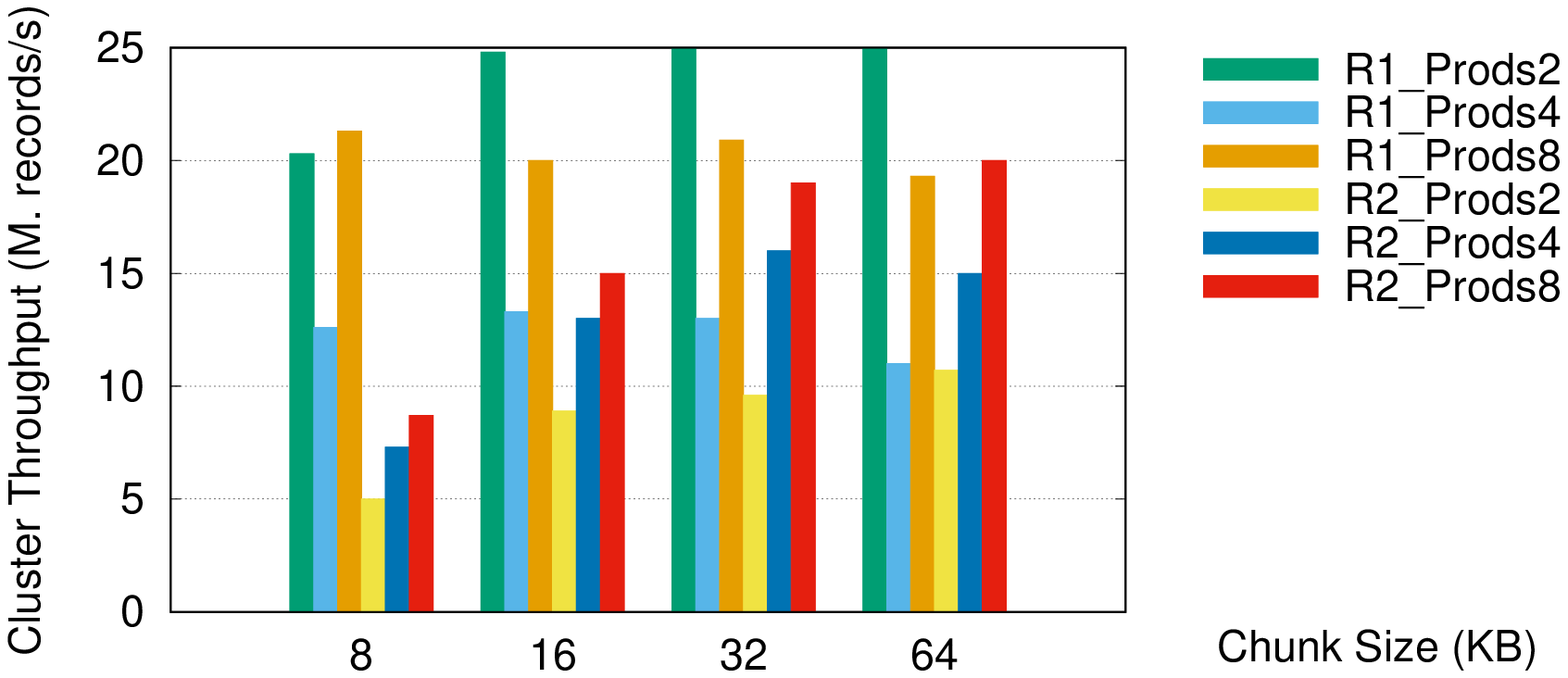}
\hspace{.03\textwidth}
\includegraphics[width=5.4cm, height=2.7cm]{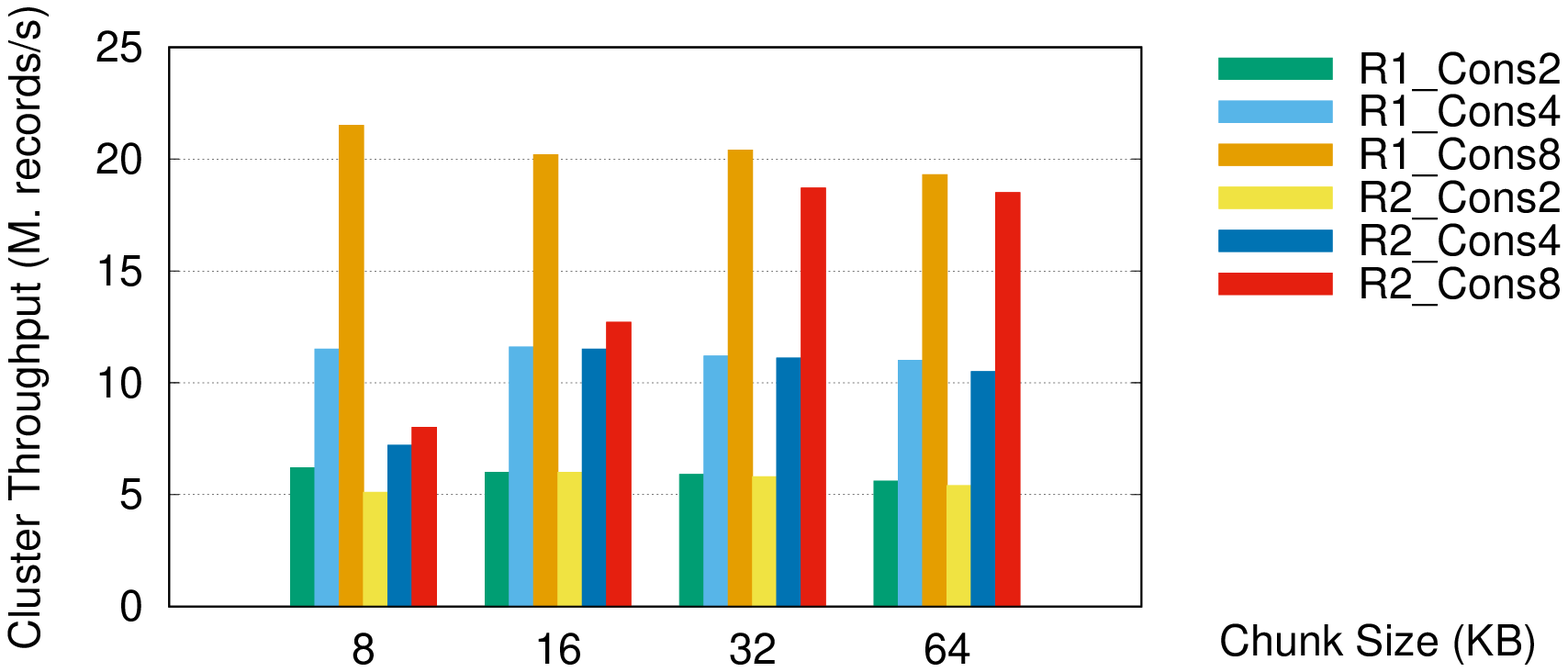}
\hspace{.03\textwidth}
\includegraphics[width=5.4cm, height=2.7cm]{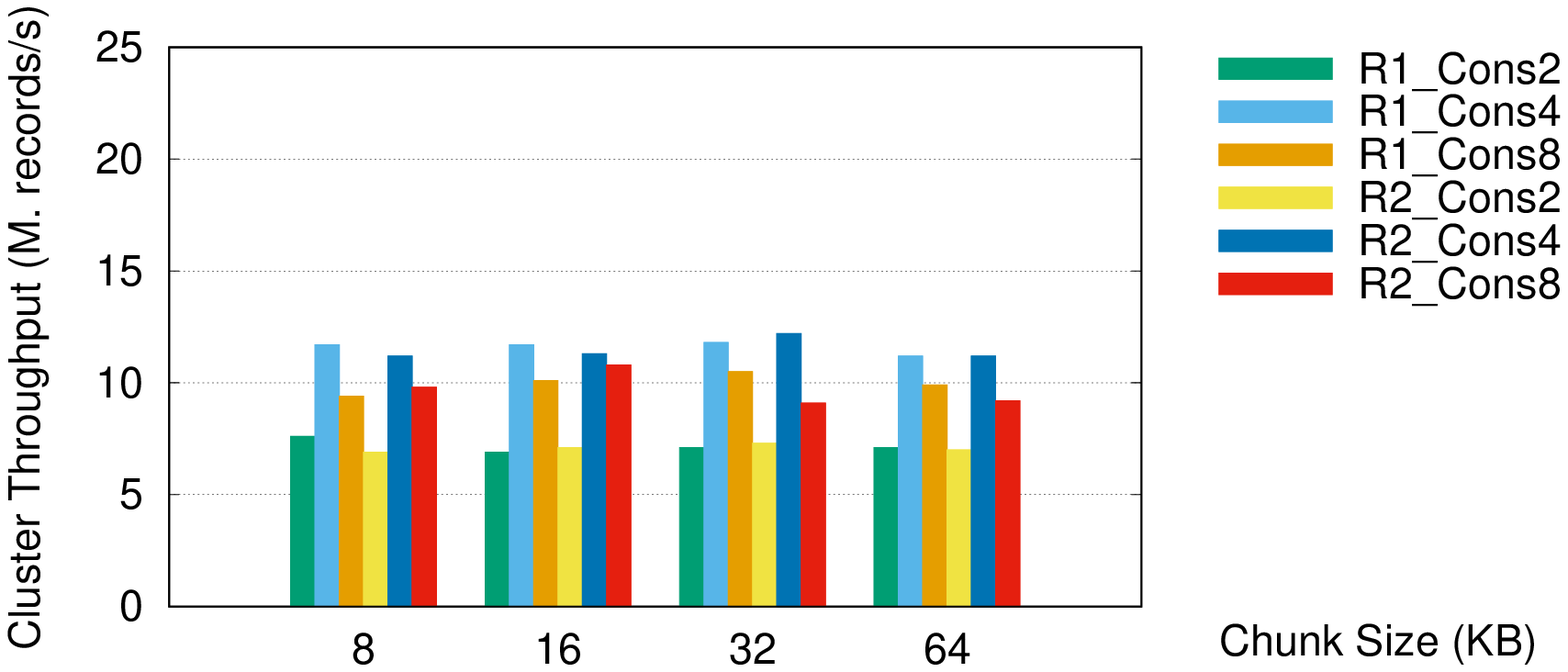}

\caption{Iterate and count benchmark for a stream with 8 partitions. Producers (left) versus pull-based consumers (middle) 
versus push-based consumers (right). R1Prods2 represent two producers with replication factor one, R2Cons8 represent eight consumers with replication factor two. Consumer chunk size is fixed to 128 KiB. We plot producer chunk size.}
\label{fig:expOnlyProdsWithCons}
\end{figure*}

\begin{figure*}[t]
\centering

\includegraphics[width=8.5cm, height=2.7cm]{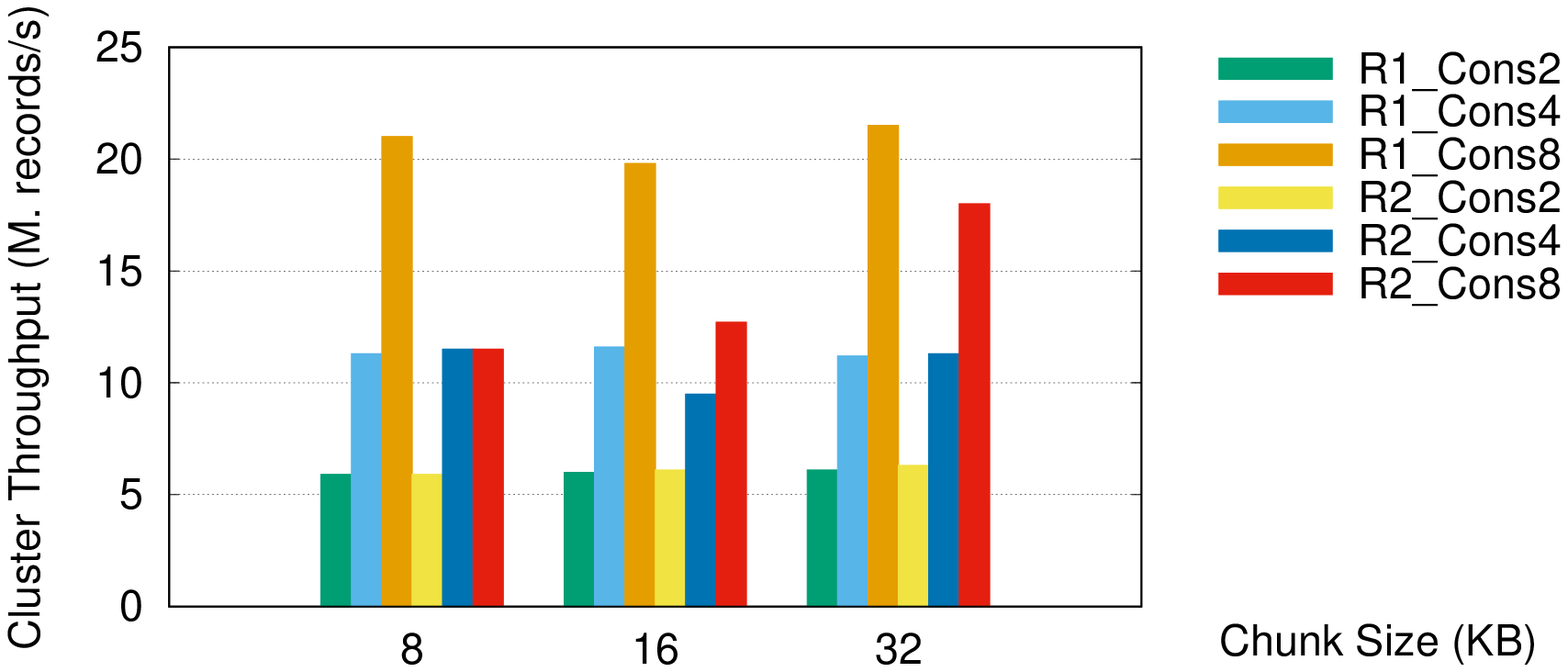}
\hspace{.03\textwidth}
\includegraphics[width=8.5cm, height=2.7cm]{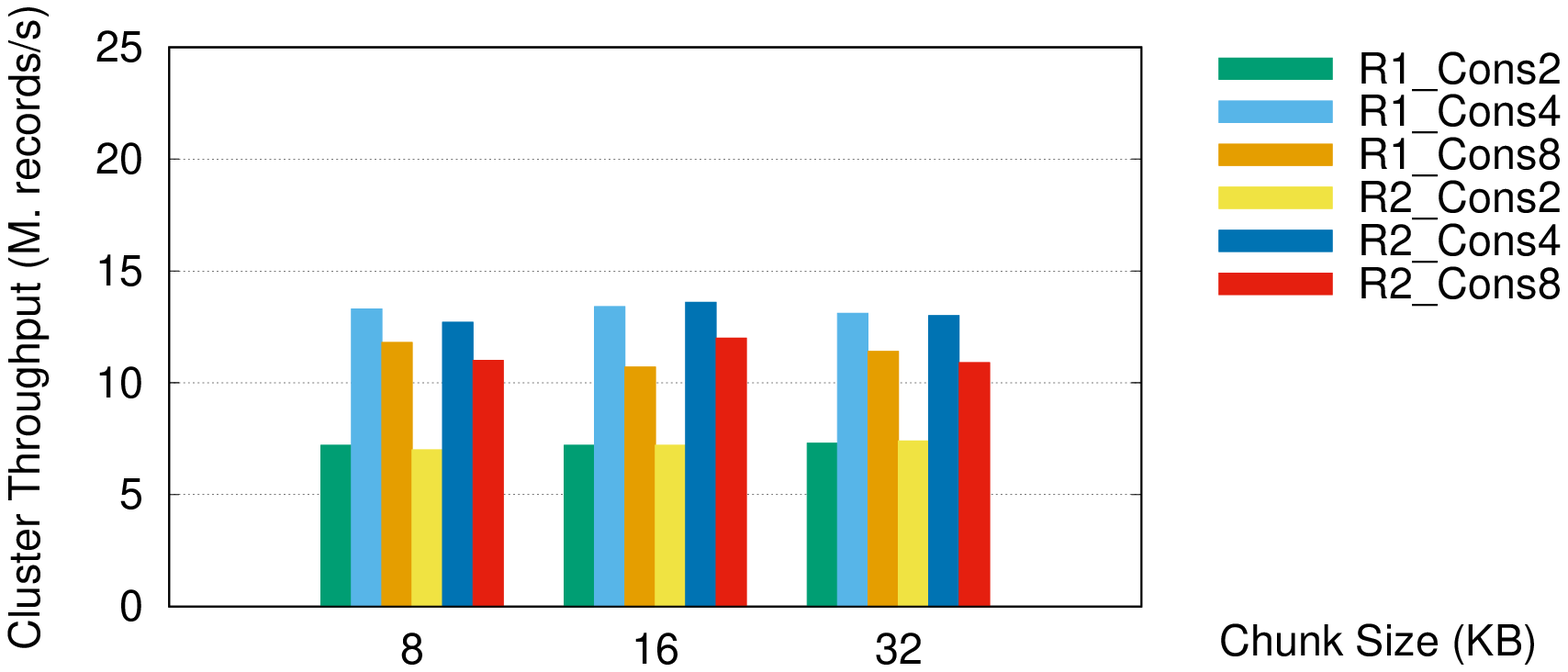}

\caption{Iterate, count and filter benchmark for a stream with 8 partitions. Pull-based consumers (left) versus push-based consumers (right). Consumer chunk size is fixed to 128 KiB. We plot producer chunk size.}
\label{fig:expOnlyProdsWithConsFilter}
\end{figure*}

\begin{figure*}[t]
\centering

\includegraphics[width=5.4cm, height=3.1cm]{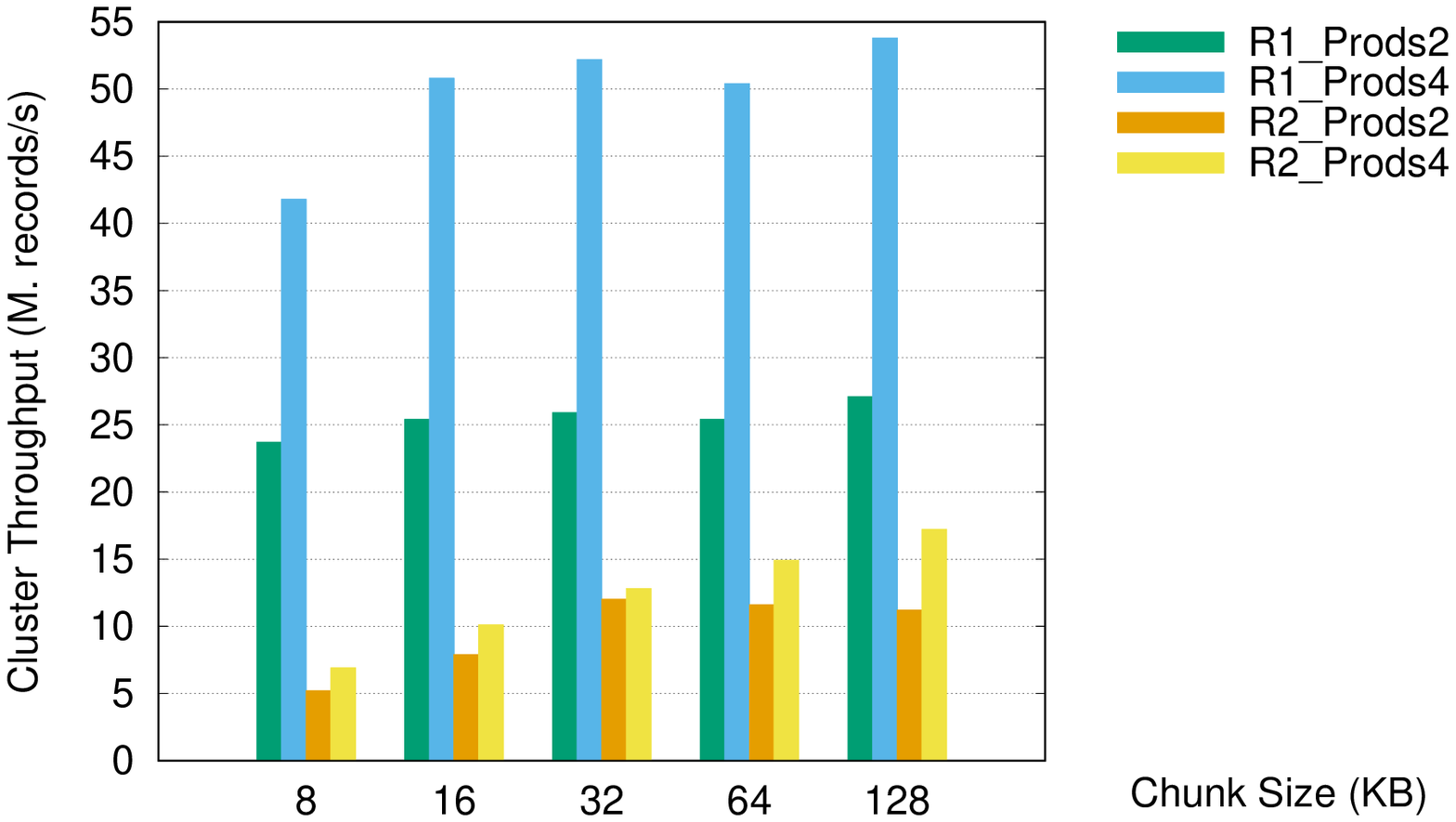}
\hspace{.03\textwidth}
\includegraphics[width=5.4cm, height=2.8cm]{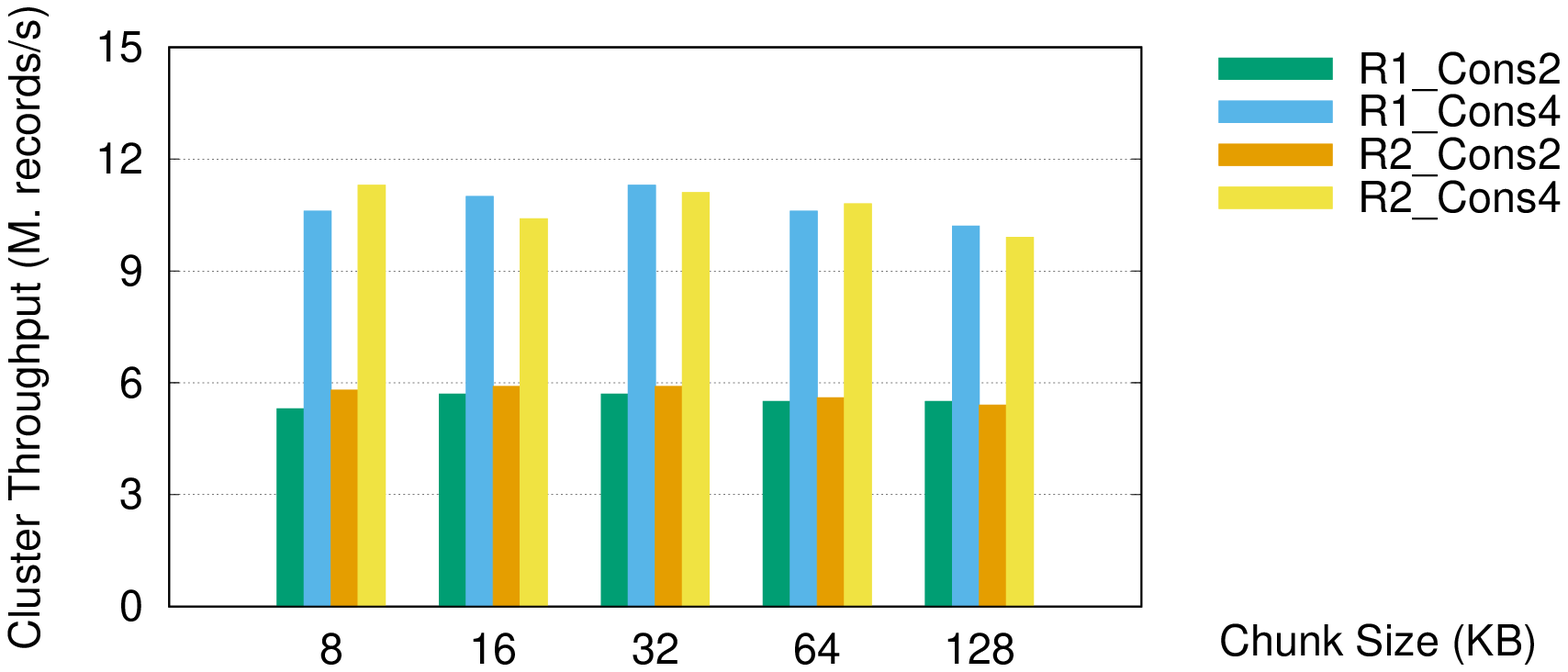}
\hspace{.03\textwidth}
\includegraphics[width=5.4cm, height=2.8cm]{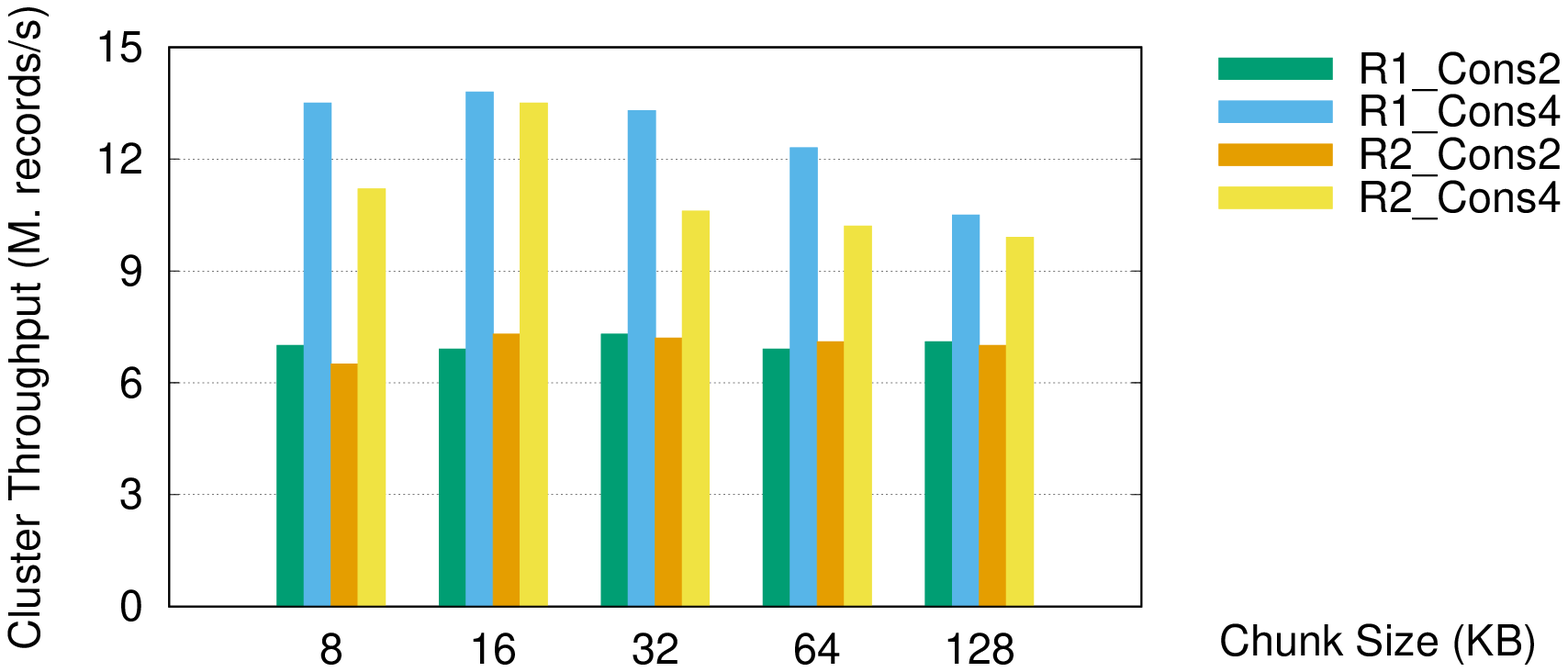}

\caption{Iterate, count and filter benchmark for a stream with 4 partitions. Producers (left) versus pull-based consumers (middle) versus push-based consumers (right). Consumer chunk size is fixed to 128 KiB. We plot producer chunk size.}
\label{fig:expOnlyProdsWithConsFilter4P}
\end{figure*}

\textbf{Synthetic benchmarks: the count operator.}
In our first evaluation, we want to understand how our chosen parameters can impact the aggregated throughput while ingesting through several producers. As illustrated in Figure \ref{fig:expOnlyProds}, we experiment with two, four, and eight concurrent producers. Increasing the chunk size \emph{CS}, the request size \emph{ReqS} increases proportionally, for a fixed record size \emph{RecS} of fixed value of 100 Bytes. While increasing the chunk size, we observe (as expected) that the cluster throughput increases; having more producers helps, although they compete at append time. We also observe that replication considerably impacts cluster throughput (as expected) since each producer has to wait for an additional replication RPC done at the broker side. Producers wait up to one millisecond before sealing chunks ready to be pushed to the broker (or the chunk gets filled and sealed) - this configuration can help trade-off throughput with latency. With only two producers, we can obtain a cluster throughput of 10 Million records per second, while we need eight producers to double this throughput. This experiment is a basis for the cluster throughput that consumers can reach for similar configurations.

\begin{figure}[t]
\centering

\includegraphics[width=8.5cm, height=3.9cm]{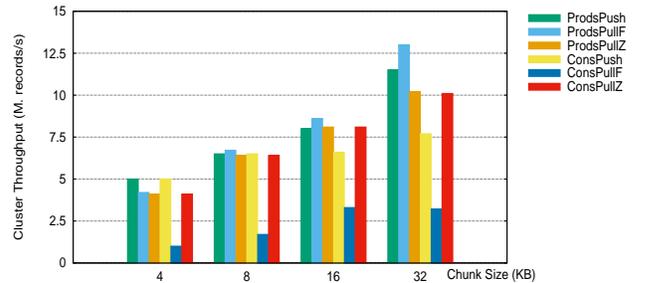}

\caption{Iterate, count and filter benchmark constrained broker resources. Comparing C++ pull-based consumers with Flink pull-based and push-based consumers. ProdsPush corresponds to producers running concurrently with push-based Flink consumers i.e. ConsPush. ProdsPullF corresponds to producers running concurrently with pull-based Flink consumers i.e. ConsPullF. ProdsPullZ corresponds to producers running concurrently with C++ pull-based consumers. Four producers and four consumers ingest and process a replicated stream (factor two) with eight partitions over one broker storage with four working cores. Consumer chunk size equals the producer chunk size.}
\label{fig:constrainedAll}
\end{figure}

\begin{figure}[t]
\centering

\includegraphics[width=8.5cm, height=3.9cm]{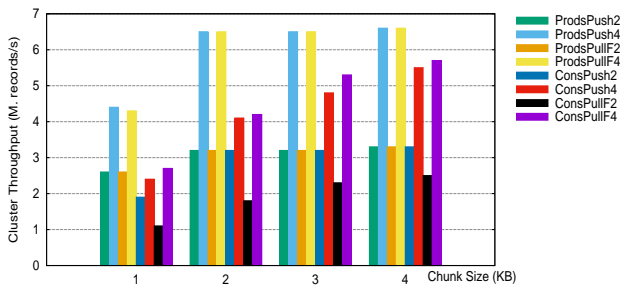}

\caption{Iterate and count benchmark stream with 8 partitions broker with 8 cores. Comparing C++ pull-based consumers with Flink pull-based and push-based consumers. ProdsPush corresponds to producers running concurrently with push-based Flink consumers i.e. ConsPush. ProdsPullF corresponds to producers running concurrently with pull-based Flink consumers i.e. ConsPullF. Consumer chunk size equals the producer chunk size multiplied by 8. We plot producer chunk size.}
\label{fig:smallChunks}
\end{figure}

\begin{figure*}[t]
\centering

\includegraphics[width=7.5cm, height=3.9cm]{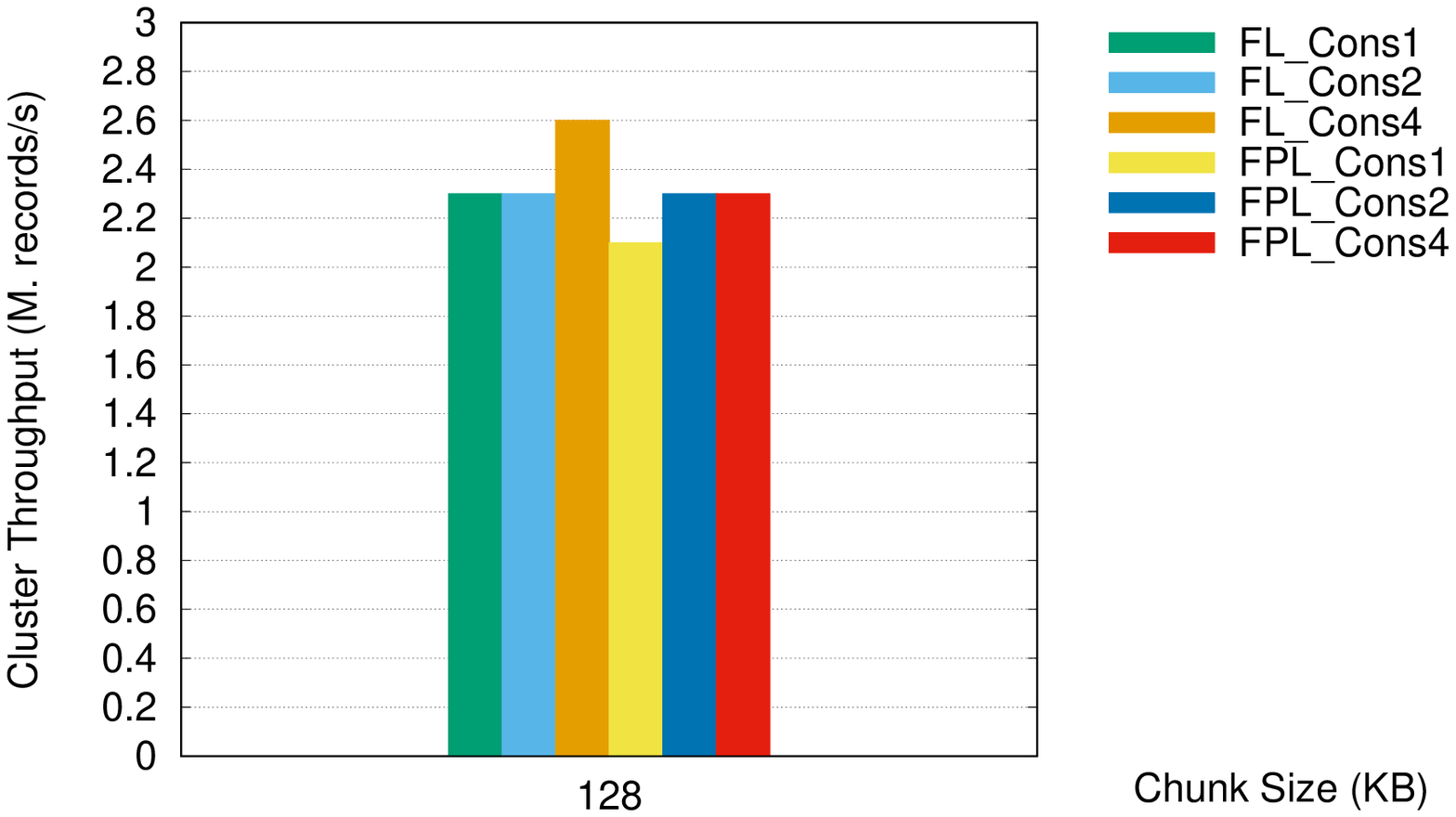}
\hspace{.03\textwidth}
\includegraphics[width=7.5cm, height=2.9cm]{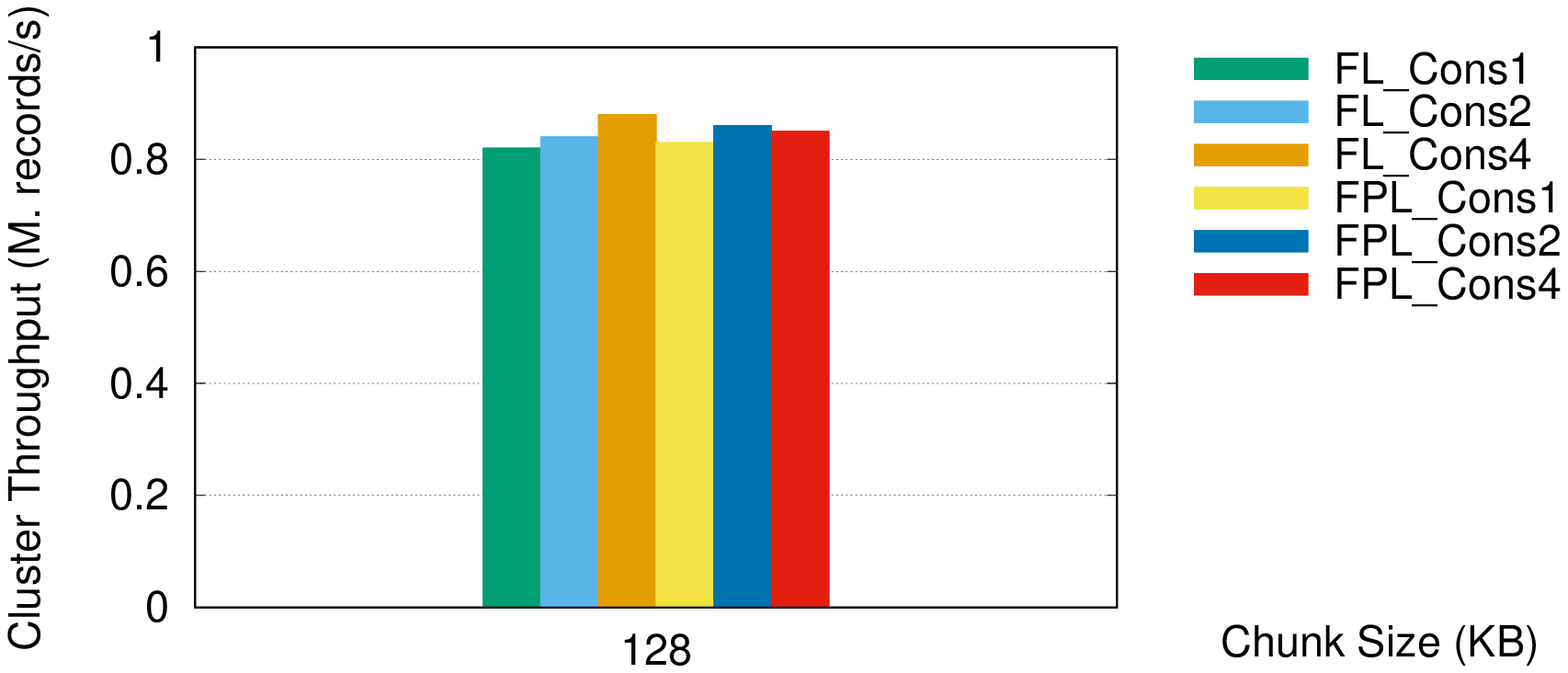}

\caption{Pull-based consumers versus push-based consumers for the word count benchmarks with 4 partitions. The left figure presents the word count benchmarks, the right figure corresponds to the windowed word count benchmark. FLCons2 represents two push-based consumers while FPLCons4 represents four pull-based consumers.}
\label{fig:expWCW4P}
\end{figure*}

%Pull-based versus push-based consumers (tuples with no key). 

The subsequent evaluation looks at concurrently running producers and consumers and compares pull-based versus push-based Flink consumers. The broker is configured with 16 working cores to accommodate up to eight producers and eight consumers concurrently writing and reading chunks of data. Since consumers compete with producers, we expect the producers' cluster throughput to drop compared to the previous evaluation that runs only concurrent producers. We show this impact in Figure \ref{fig:expOnlyProdsWithCons}: due to higher competition to broker resources by consumers, producers obtain a reduced cluster throughput compared to the previous experiment. The number of consumers similarly limits the consumers' cluster throughput. However, in most configurations, consumers fail to keep up with the producers' rate. 

When comparing pull-based with push-based consumers, we first observe that the configuration with eight consumers does not scale in the push-based strategy due to the limitations of the dedicated thread pushing the following chunks of data. However, (although with eight consumers the pull-based strategy can obtain a better cluster throughput,) for up to four consumers the push-based strategy not only can obtain slightly better cluster throughput but the number of resources dedicated to consumers reduces considerably (two threads versus eight threads for the configuration with four consumers).
While pull-based consumers double the cluster throughput when using 16 threads for the source operators, push-based consumers only use two threads for the source operator.

\textbf{Synthetic Benchmarks: The Filter Operator.} We further compare pull-based versus push-based consumers when implementing the filter operator, in addition, to counting for a stream with eight partitions. Similar to previous experiments, the push-based consumers are slower when scaled to eight for larger chunks, as illustrated in Figure \ref{fig:expOnlyProdsWithConsFilter}.

%next exp1608FilterCount 4 partitions, CS 8 16 32 64 128 KB

As illustrated in Figure \ref{fig:expOnlyProdsWithConsFilter4P}, when experimenting with up to four producers and four consumers over a stream with four partitions, the push-based strategy provides a cluster throughput slightly higher with smaller chunks, being able to process two million tuples per second additionally over the pull-based approach. With larger chunks, the throughput reduces: architects have to carrefully tune the chunk size in order to get the best performance.

When experimenting with smaller chunks (producers' chunk size is one to four KiB, consumers get 8x higher chunks to try to keep up with producers), more work needs to be done by pull-based consumers since they have to issue more frequently RPCs (see Figure \ref{fig:smallChunks}). Moreover, the push-based strategy provides higher or similar cluster throughput than the pull-based strategy while using fewer resources.

Next, we design an experiment with constrained resources for the storage and backup brokers configured with four cores. We ingest data from four producers into a replicated stream (factor two) with eight partitions. We concurrently run four consumers configured to use Flink-based push and pull strategies and native C++ pull-based consumers. Consumers iterate, filter and count tuples that are reported every second by eight mappers. We report our results in Figure \ref{fig:constrainedAll} where we compare the cluster throughput of both producers and consumers. Producers compete directly with pull-based consumers, and we expect the cluster throughput to be higher when concurrent consumers use a push-based strategy. However, producers' results are similar except for the 32 KB chunk size when producers manage to push more data since pull-based consumers are slower. We observe that the C++ pull-based consumers can better keep up with producers while push-based consumers can keep up with producers when configured to use smaller chunks. The push-based strategy for Flink is up to 2x better than the pull-based strategy of Flink consumers. Consequently, the push-based approach can be more performant for resource-constrained scenarios.

\textbf{Wikipedia Benchmarks: (Windowed) Word Count Streaming.}
For the following experiments, the producers are configured to read Wikipedia files in chunks with records of 2 KiB. Therefore, producers can push about 2 GiB of text in a few seconds. Consumers run for tens of seconds and do not compete with producers. As illustrated in Figure \ref{fig:expWCW4P}, pull-based and push-based consumers demonstrate similar performance. We plot word count tuples per second aggregated for eight mappers while scaling consumers from one to four. Although not shown, results are similar when we experiment with smaller chunks or streams with more partitions since this benchmark is CPU-bound. To avoid network bottlenecks when processing large datasets like this one (e.g., tens of GBs) on commodity clusters, the push-based approach can be more competitive when pushing pre-processing and local aggregations at the storage.

\section{Discussion and Future Implementation Optimizations}

Layered storage and processing Fast Data architectures decouple storage and processing engines to give more flexibility to architects looking to explore various frameworks for responding to different application needs. These architectures implement pull-based (RPC) consumers to separate application workload from storage usage. Therefore, components of layered streaming architectures can more easily scale independently by adding more storage or processing nodes as needed. 

As proposed and evaluated in this paper, we observe that a push-based strategy can improve performance for high-throughput and low-latency scenarios. However, for applications that can estimate workloads and overprovision cluster storage/processing resources, a pull-based approach for streaming consumers can be enough, avoiding more complex architectures like the one we propose. Moreover, architects should highly consider unpredictable workloads that can overload storage resources, leading to low performance, while also considering optimizing resource usage. In this case, a push-based strategy could be worth the deployment and development efforts, potentially providing similar or better throughput to a pull-based approach while reducing resources required by low-latency and high-throughput consumers.

Regarding our prototype implementation, we believe there is room for further improvements. One future step is integrating the shared object store and notifications mechanism inside the broker storage implementation. This choice will bring up two potential optimizations. Firstly, it would allow avoiding another copy of data by leveraging existing in-memory segments that store partition data. Secondly, we could optimize latency by implementing the notification mechanism through the asynchronous RPCs available in KerA or RDMA when consumers are deployed separately from storage. Furthermore, applying pre-processing functions directly at the storage engine (e.g., as done in \cite{inmemorystorage}) reduces the necessary data to be pushed and avoids initial serialization done in the streaming engine.

\section{Conclusion}

We have proposed a unified real-time storage and processing architecture that leverages a push-based strategy for streaming consumers. Experimental evaluations show that when storage resources are enough for concurrent producers and consumers, the push-based approach is performance competitive with the pull-based one (as currently implemented in state-of-the-art real-time architectures) while consuming fewer resources. However, when the competition of concurrent producers and consumers intensifies and the storage resources (i.e., number of cores) are more constrained, the push-based strategy can enable a better throughput by a factor of up to 2x while reducing processing latency.

Furthermore, given that we experiment on high-end hardware, we believe that the push-based streaming strategy is even more competitive when deployed on commodity hardware for both low-latency and high-throughput scenarios. Our next step is to leverage consumer offsets when implementing a unified real-time architecture and to explore fast crash recovery techniques for real-time storage and processing deployed on multi-core nodes and low-latency networking. We believe that by adopting a granular/composable architectural approach, a unified real-time storage and processing engine could provide millisecond recovery time while maintaining properties like durability and exactly-once processing. Finally, we are looking to propose unified storage and real-time processing model that can help developers by automatically estimating and deploying optimized configurations that can employ pull-based and push-based streaming strategies.

%%%%%%%%%%%%%%%%%%%%%%%%%%%%%%%%%%%%%%%%%%%%%%%%%%%%%%%%%%%%%%
\section*{Acknowledgment} The experiments presented in this paper were carried
out using the HPC facilities of the University of Luxembourg~\cite{uniluhpc} -- see hpc.uni.lu. This work is done in the context of bridging clouds and supercomputers, a project in collaboration with LuxProvide.

\bibliographystyle{IEEEtran}

\bibliography{sloc}

\end{document}